\documentclass[]{aiaa-tc}% insert '[draft]' option to show overfull boxes
\usepackage{booktabs,amsmath,amsthm,amssymb,amsbsy,latexsym,graphicx,color}
\usepackage{rotating}
\usepackage{float}
\usepackage{subfigure}
\usepackage{upgreek}
\usepackage{bm}
\usepackage{mathdots}
\usepackage{relsize}
\usepackage{caption}
\usepackage{newfloat}
\usepackage{fixltx2e}
\usepackage{nomencl}
\usepackage{multicol}
\usepackage{indentfirst}
\usepackage{etoolbox}
\usepackage{tocvsec2}
\usepackage[titletoc]{appendix}
\usepackage{appendix}
\usepackage[bbgreekl]{mathbbol}
\usepackage{graphicx}

\title{A Finite Strain Constitutive Model for Martensitic Transformation in Shape Memory Alloys Based on Logarithmic Strain}

 \author
  {
  	Lei Xu
  	\thanks{Graduate Research Assistant, Aerospace Engineering, College Station, TX 77843, USA, Student member.}\thanksibid{1}\\
  {\normalsize\itshape
   Texas A\&M University, College Station, TX 77843, USA}\\
  \and
  Theocharis Baxevanis
  \thanks{Assistant Professor,Mechanical Engineering,  Houston, TX 77204, USA, Faculty member.}\thanksibid{2}\\
  {\normalsize\itshape
	 University of Houston, Houston, TX 77204, USA}\\
  \and
   Dimitris Lagoudas
   \thanks{ Distinguished Professor,Aerospace Engineering,College Station, TX 77843, USA, Faculty member.}\thanksibid{1}\\
  {\normalsize\itshape
  Texas A\&M University, College Station, TX 77843, USA}\\
  }

 \AIAApapernumber{2016-NUMBER}
 \AIAAconference{AIAA SciTech, 9–13 January 2017, Gaylord Texan, Grapevine, Texas}
 \AIAAcopyright{\AIAAcopyrightD{2016}}

\begin{document}

\maketitle

\begin{abstract}
Shape Memory Alloys (SMAs) are materials with the ability to recover apparently permanent deformation under specific thermomechanical loading. The majority of constitutive models for SMAs are developed based on the infinitesimal strain theory. However, such assumption may not be proper in the presence of geometric discontinuities, such as cracks, and repeated cycling loading that has been reported to induce irrecoverable strains up to 20\% due to transformation induced plasticity. In addition to finite strains, SMA-based devices may also undergo large rotations. Thus, it is indispensable to develop a constitutive model based on the finite strain to provide accurate predictions of these actuators response. A three-dimensional phenomenological constitutive model for SMAs considering finite strains and finite rotations is proposed in this work. This model utilizes the logarithmic strain as the strain measure that is the strain measure whose logarithmic rate in a corotating material frame is equal to the rate of deformation tensor. In the proposed model, the martensitic volume fraction and the second-order logarithmic transformation strain tensor are chosen as the internal state variables associated with the inelastic transformation process. Numerical simulations considering basic SMAs component geometries such as a bar, a beam, and a torque tube are performed to test the capabilities of the proposed model under both mechanically and thermally induced phase transformation. For numerical examples in which the SMA components exhibits finite strains along with finite rotations, discrepancies are observed between the responses predicted by the present model and its infinitesimal counterpart. Also, the spurious accumulated residual stress observed in infinitesimal strain model is eliminated by the proposed model. This shows that the infinitesimal strain assumption is not applicable in such cases and the proposed model considering large strains and rotations is needed to provide accurate predictions. The presented model formulation will be extended in future work for the incorporation of transformation-induced plasticity.
\end{abstract}

\newpage
\section*{Nomenclature}

\begin{multicols}{2}
\begin{tabbing}
  \bfseries left \quad \=\bfseries center \quad \=\bfseries right \quad \=\bfseries paragraph \kill% this line sets tab stop
  $\mathbf{B}$ \> \quad Left Cauchy-Green tensor \\ 
  $\mathbf{B}_{i,j}$ \> \quad Subordinate eigenprojections of $\mathbf{B}$  \\  
  $\mathbf{D}$ \>\quad Rate of deformation tensor \\
  $\mathbf{D}^e$ \>\quad Elastic rate of deformation tensor \\
  $\mathbf{D}^{tr}$ \>\quad Dissipative rate of deformation tensor \\
  $\mathbf{F}$ \>\quad Deformation gradient \\
  $\mathbf{F}^e$ \>\quad Elastic deformation gradient \\
  $\mathbf{F}^{tr}$ \>\quad Dissipative deformation gradient \\
  $\mathbf{L}$ \>\quad Velocity gradient\\  
  $\mathbf{S}$ \>\quad Forth order compliance tensor\\  
  $\mathbf{S}^A$ \>\quad Forth order compliance tensor of austenite\\  
  $\mathbf{S}^M$ \>\quad Forth order compliance tensor of martensite\\ 
  $\Delta \mathbf{S}$ \>\quad Difference of compliance tensor between\\
				 \>\quad austenite and martensite \\   
  $\mathbf{W}$ \>\quad Anti-symmetric part of velocity gradient\\
  $\mathbf{X}$ \>\quad Position vector in reference configuration\\
  $\bm\Lambda$ \>\quad Transformation direction tensor\\
  $\bm\Lambda^{fwd}$ \>\quad Forward transformation direction tensor\\
  $\bm\Lambda^{rev}$ \>\quad Reverse transformation direction tensor\\
  $\mathbf{\Omega}^{log}$ \>\quad Logarithmic spin\\
  $\mathbf{h}$ \>\quad Logarithmic strain of Eulerian type \\
  $\mathbf{h}^{tr}$ \>\quad Transformation logarithmic strain \\ 
  $\mathbf{h}^{tr-r}$ \>\quad Transformation logarithmic strain at  \\ 
				  	   \>\quad reverse point \\  
  $\bar{\mathbf{h}}^{tr-r}$ \>\quad Effective transformation strain at reverse \\
						   \>\quad point \\      
  $\mathbf{x}$ \>\quad Position vector at current configuration\\

  $\bm{\Upsilon}$ \>\quad Set of internal state variables\\
  $\bm{\uptau}$ \>\quad Kirchhoff stress \\ 
  $\bm{\uptau}'$ \>\quad Deviatoric part of kirchhoff stress \\ 
  ${\bar\uptau}^{'}$ \>\quad Effective kirchhoff stress \\   
  $\bm\alpha^A$ \>\quad Second order thermal expansion for austenite   \\
  $\bm\alpha^M$ \>\quad Second order thermal expansion for martensite  \\  
  $\Delta\alpha$ \>\quad Difference of thermal expansion \\      
  $\mathcal{D} $ \>\quad Dissipation energy\\   
  
  $A_s$ \>\quad Austenite transformation start temperature\\
  $A_f$ \>\quad Austenite transformation finish temperature\\  
  $M_s$ \>\quad Martensite transformation start temperature\\
  $M_f$ \>\quad Martensite transformation finish temperature\\       
  $G$ \>\quad Gibbs free energy\\
  $H^{max}$ \>\quad Maximum transformation strain\\  
  $T$ \>\quad Temperature \\ 
  $T_0$ \>\quad Temperature at reference point \\ 
  $Y$ \>\quad Critical thermodynamic driving force \\   

  $b_1,b_2$ \>\quad Material parameters in hardening function \\ 
  $c$ \>\quad Specific heat \\       
  $\mu_1,\mu_2$ \>\quad Material parameters in hardening function \\       
  $f(\xi)$ \>\quad Hardening function \\     
  $s$ \>\quad Specific entropy \\
  $s_0$ \>\quad Specific entropy at reference state \\  
  $\Delta s_0$ \>\quad Difference of specific entropy \\    
  $u$ \>\quad Internal energy\\
  $u_0$ \>\quad Internal energy at reference state\\  
  $\Delta u_0$ \>\quad Difference of internal energy \\

  $\Phi$ \>\quad Transformation function\\
  $\rho$ \>\quad Density at current configuration  \\
  $\rho_{0}$ \>\quad Density at reference configuration  \\
  $\xi$ \>\quad Martensite volume fraction  \\
  $\nabla$ \>\quad Gradient operator\\
  $\chi$ \>\quad Deformation mapping function\\
  $\lambda_i,\lambda_j$ \>\quad Eigenvalues of $\mathbf{B}$  \\
  $\pi$ \>\quad Thermodynamic driving force\\

\end{tabbing}
\end{multicols}

\section{Introduction}

Shape memory alloys (SMAs) are a kind of specific materials with the ability to recover its pre-defined shape under thermal-mechanical loadings. Since the discovery of shape memory effect phenomenon among metallic alloys from 1930s to 1950s (Otsuka and Wayman, 1932; Greninger and Mooradian, 1938; Kurdjumov and Khandros, 1949; Chang and Read, 1951), shape memory alloys has been extensively investigated to be used as sensors, controllers and actuators etc. towards building smart system integrated with adaptive and intelligent functions.

Over the several past decades, a substantial of SMAs constitutive theories at continuum levels have been proposed, the majority of them are within small deformation regime based on infinitesimal strain assumption. Some thorough review of shape memory models can be found from Boyd and Lagoudas\cite{boyd1996}, Birman and November\cite{birman1997}, Raniecki and Lexcellent \cite{raniecki1992,raniecki1998}, Patoor et al.\cite{patoor2006,patoor1996}, Hackl and Heinen\cite{hackl2008}, Levitas and Preston \cite{levitas1998,levitas2002} etc. In general, these models can be categorized into three different types: crystal-plasticity based model, phase field method based model and phenomenological plasticity based models. In crystal-plasticity based model, it follows the multiplicative decomposition of deformation gradient $\mathbf{F}$ into a recoverable part $ \mathbf{F^{e}} $ multiplied by an inelastic dissipative transformation part $\mathbf{F^{tr}}$. One main merit of this model type is it takes into account the crystal orientation, hence it can capture the tension compression asymmetry phenomenon exhibited by many experiment SMAs samples. This kind of model are more physical related from microstructure point of view. However, the complex implementation process of this model type makes it highly computational costly. Some example on crystal-plasticity based models can be found from Auricchio and Taylor\cite{auricchio1997}, Thamburaja and Anand\cite{tham2001}, Wang at al.\cite{Wang2008}, Reese and Christ\cite{reese2008}, Yu et al.\cite{yu2013}. Another method to consider microstructure evolution is phase field method based approach. The key point of this method is to utilize an order parameter to differentiate different phases in SMAs, through which it can keep tracking the microstructure  changing (like phase boundary) during SMAs transformation process , which makes it particularly suited to studying the dynamic evolution of martensitic microstructures. Some pioneering work related to phase field method can be obtained from Levitas et al.\cite{levitas1998,levitas2002}, Chen et al.\cite{chen2002}, Steinbach et al.\cite{steinbach1999,steinbach2006}, Mamivand et al.\cite{mamivand2013}, Zhong and Zhu\cite{zhong2014}. On the other hand, phenomenological plasticity based approach is following the legacy of phenomenological $J_2$ plasticity theory, it starts from an additive decomposition of total strain into an elastic part plus inelastic or transformation part based on infinitesimal strain assumption, additionally it introduces internal state variables (such as phase volume fraction) to capture the response of bulk material in a macroscopic way. Although it loses microscopic information on microstructure, simplicity of these models and its well established computational implementation procedure makes it widely used in design for SMAs components  among engineering field, especially for complex SMAs structure with multiaxial loading conditions. Published and well accepted example models falling into this type can be obtained from literature Lagoudas et al. \cite{lagoudas2008,lagoudas2012}, Brinson et al.\cite{brinson1993finite,brinson1993one}, Lexcellent et al. \cite{lexcellent1996general,lexcellent2013shape}.

When the deformation regime is within infinitesimal strain range, all the above mentioned models are able to predict material response accurately. However, recent publication has reported that shape memory alloys can reversibly deform to a relatively large strain up to 8\% \cite{jani2014,shaw2000}, and also repeated cycling loading has been reported to induce irrecoverable transformation induced plasticity strains up to 20\% . In addition to such relatively large strains, SMAs-based devices may also undergo finite rotations during its deployment. Combining all the above factors, it is indispensable to develop a constitutive model based on finite deformation framework to provide accurate predictions of these actuators response when deformed. As for the SMA model at the frame work of finite deformation, crystal-plasticity based models utilizing multiplicative decomposition is built within finite-deformation configuration, but again, the implementation complexity of the models hinders its attractiveness for application design. Recent constitutive theories of this model type can be found from literature Auricchio\cite{auricchio2001}, Ziolkowski \cite{ziolkowski2007}, Christ and Reese\cite{christ2009}, Reese and Christ\cite{reese2008}, Evangelista et al.\cite{evangelista2010}, Arghavani and Auricchio\cite{arghavani2011}. On the other hand, phenomenological $J_2$ plasticity based models building on infinitesimal strain assumption, though it runs much faster in numerical simulation, may not be proper in the presence of such large strains. Much effort has been devoted to extend this type of model to be used in finite strain deformation analysis. One way is to set up a direct relation between  the rate of deformation tensor and an objective rate on a finite strain measure. Utilize the additive decomposition of the rate of deformation tensor $\mathbf{D}$ into an elastic part $\mathbf{D}^e$ plus a dissipative part $\mathbf{D}^{tr}$. This approach requires to adopt an objective rate to achieve the principle of objectivity for the rate form hypo-elastic constitutive equation. Well-known existing objective rates such as Zaremba-Jaumann-Noll rate, Green-Naghdi-Dienes rate, Truesdell rate etc. have been proposed to achieve such goal. However, above mentioned objective rates are not real 'objective', they fail to be integrated from rate form hypo-elastic equation to yield a recoverable hyper-elastic equation. Namely, a non-integrable hypo-elastic formulation is path-dependent and dissipative, and thus would deviate essentially from the recoverable elastic-like behavior\cite{xiao2006}. It was not until the logarithmic rate proposed by Xiao et al.\cite{xiao1997,xiao1997hypo}, Bruhns et al.\cite{bruhns1999self,bruhns2001large,bruhns2001self}, Meyers et al.\cite{meyers2003elastic,meyers2006choice} that the non-integral issue of objective rates has been resolved.

Moreover, we know a proper finite strain measure is very important for finite deformation analysis. In this paper, we are going to use the logarithmic stain as the finite strain measure, as a result of following reasons: (1)It has been proved by Xiao et al.\cite{xiao1997} that the logarithmic rate of logarithmic strain $ \mathbf{h}$ is exactly identical with the rate of deformation tensor $\mathbf{D}$, and logarithmic strain is the only one among all other strain measures enjoying this important property, which can be utilized in the thermodynamic framework to make the derivation of constitutive equation in a fully consistent way. (2)Because of the mathematical property of natural logarithm, the total logarithmic strain can be additively decomposed into volumetric part and deviatoric part, while those two portions are inevitably coupled at all other strain measures, such as Green-Lagrange train, used in finite deformation analysis.

Impressed by the above facts, a finite strain constitutive model based on logarithmic strain to analyze the martensitic transformation for shape memory alloys is going to be proposed in this article. The model is based on the SMAs model proposed by Lagoudas and coworkers \cite{boyd1996,lagoudas2008,lagoudas2012} for small deformation case. To this end, the paper is organized as follows. In Section~\ref{Preliminary}, we represent some preliminaries on kinematics in continuum mechanics. Section~\ref{Model} will concentrate on the thermodymamics framework to formulate the SMA model by using logarithmic strain and logarithmic rate. Boundary value problems will be addressed to test the capability of proposed model in Section~\ref{Result}. At the end, we summarize this paper with conclusion in Section~\ref{Conc}

%In Section~\ref{Implementation}, the numerical implementation of the finite strain model is presented. Boundary value problems will be addressed to test the capability of proposed model in Section~\ref{Result}. At the end, we summarize this paper with conclusion in Section~\ref{Conc} .   

%%==========================================================================
\section{Preliminaries} \label{Preliminary}
\subsection{Kinematics}
Let body $ \mathcal{B} $ with its material points defined by position vector $ \mathbf{X} $ in the reference (undeformed) configuration at initial time $ t_{0} $, and let vector $ \mathbf{x} $ represent the position vector occupied by material points $ \mathbf{X} $  after deformation at current (deformed) configuration at time $ t $, the mapping is defined by $ \mathbf{x=\chi}(\mathbf{X},t) $. The deformation process from the initial configuration to current configuration can be characterized by the deformation gradient $\mathbf{F}(\mathbf{x},t)$:
\begin{equation}\label{Deformation}
\mathbf{F}(\mathbf{x},t) =\frac{\partial \mathbf{x}}{ \partial \mathbf{X}}  
\end{equation}
Then, the velocity gradient $\mathbf{L}$ is defined through as follows:
\begin{equation}\label{Vel_g}
\mathbf{L} =\mathbf{\dot{F}}\mathbf{F} ^{-1} 
\end{equation}
Velocity gradient $\mathbf{L}$ can be additively decomposed into a symmetric part called the rate of deformation tensor, i.e. $\mathbf{D}$, plus an anti-symmetric part called the spin tensor, i.e. $\mathbf{W}$. 
\begin{equation}\label{S_P_tensor}
\begin{array}{c}
\mathbf{L = D + W}; \quad
\begin{cases} \mathbf{D} =\dfrac{1}{2} \mathbf{(L+L^{T})}, \vspace{5pt}  \\ 
\mathbf{W} =\dfrac{1}{2} \mathbf{(L-L^{T})}, \\  
\end{cases}
\end{array}
\end{equation}
The following polar decomposition formula is well known, in which $\mathbf{R}$ is the rotation tensor and $\mathbf{V}$ is the left stretch.
\begin{equation}\label{Polar_Dec}
\mathbf{F = VR}
\end{equation}
The left Cauchy-Green tensor $\mathbf{B}$ is defined by
\begin{equation}\label{LCG_tensor}
\mathbf{B} = \mathbf{FF}^{T} =\mathbf{V}^2 
\end{equation}
The logarithmic strain of Eulerian type $ \mathbf{h} $ is given through,
\begin{equation}\label{Log_strain}
\mathbf{h} = \frac{1}{2} \ln\mathbf{ {B}} =\mathbf{\ln {V}}
\end{equation}

\subsection{Logarithmic rate and Logarithmic spin}\label{Logarithmic}
In finite elastoplasticity theory, the additive decomposition of the total rate of deformation tensor $\mathbf{D}$ into an elastic part $\mathbf{D}^e$ plus a dissipative part $\mathbf{D}^{tr}$ was successfully applied in finite deformation analysis. One of the main job in it is to adopt an appropriate objective rates to achieve the principle of objectivity in rate form equations. Many objective rates have been proposed by different scholars. However, none of them was able to set up a direct relation between the rate of deformation tensor $ \mathbf{D} $ and an objective rate of strain measure, thus many spurious phenomenons, such as shear stress oscillation, dissipative energy or residual stress accumulated in elastic deformation etc., are observed. Until recently the so-called logarithmic rate proposed by Xiao et al. \cite{xiao1997,xiao1997hypo,xiao2006}, Bruhns et al.\cite{bruhns1999self,bruhns2001large,bruhns2001self}, Meyers et al.\cite{meyers2003elastic,meyers2006choice} 
successfully resolved such self-inconsistent issues. As they showed that the logarithmic rate of the logarithmic strain $\mathbf{h}$ of Eulerian type is identical with the rate of deformation tensor $\mathbf{D}$, which is expressed as:
\begin{equation}\label{eq:Log_strain_rate}
\mathring{\mathbf{h}}^{log} = \dot{\mathbf{h}}+\mathbf{h} \mathbf{ \Omega}^{log}-\mathbf{ \Omega}^{log}\mathbf{h}= \mathbf{D}
\end{equation}
Where $\mathring{\mathbf{h}}^{log}$ means the logarithmic rate of logarithmic strain and $\dot{\mathbf{h}}$ is the conventional time rate of logarithmic strain.  $ \mathbf{ \Omega}^{log} $ is called logarithmic spin introduced by Xiao and Bruhns \cite{xiao1997} with explicit expression as:
\begin{equation}\label{eq:Log_spin}
\mathbf{\Omega}^{log} = \mathbf{W}+ \sum_{i \neq j}^{n}  \big(\frac{1+(\lambda_{i}/\lambda_{j})}{1-(\lambda_{i}/\lambda_{j})}+\frac{2}{•\ln (\lambda_{i}/\lambda_{j})}\big) \mathbf{B}_i \mathbf{D} \mathbf{B}_j
\end{equation}
Where $ \mathbf{W} $ is the spin tensor; $\lambda_{i,j} (i,j=1,2,3...) $ are the eigenvalues of Left Cauchy-Green tensors $ \mathbf{B} $; $ \mathbf{B}_{i,j} $ are the corresponding subordinate eigenprojections of $ \mathbf{B}$. Equation \ref{eq:Log_strain_rate} is very important in consistent formulation of finite strain SMAs model at following thermodynamic framework later on . 

\subsection{Additive decomposition of logarithmic strain }\label{AdditiveStrain}
Starting from additive decomposition of $\mathbf{D}$ in kinematics for deformation,
\begin{equation}\label{eq:add_D}
\mathbf{D}=\mathbf{D}^{e}+\mathbf{D}^{tr}
\end{equation}
The total stress power supplied from outside working on body $ \mathcal{B} $ per unit volume can be calculated and additively decomposed into,
\begin{equation}\label{eq:power}
P_{stress\_power}=\bm{\uptau}:\mathbf{D}=\bm{\uptau}:\mathbf{D}^{e}+\bm{\uptau}:\mathbf{D}^{tr}
\end{equation}
From energy point of view, additive decomposition in deformation kinematics can be interpreted as total stress power being split into a recoverable part as $\bm{\uptau}:\mathbf{D}^{e}$ plus an irrecoverable part as $\bm{\uptau}:\mathbf{D}^{tr}$ associated with dissipative process (such as plasticity deformation, transformation process etc.).

By virtue of equation \ref{eq:Log_strain_rate}, elastic part $\mathbf{D}^{e}$ and dissipative part  $\mathbf{D}^{tr}$ in equation \ref{eq:add_D} can be rewritten as the logarithmic rate of elastic logarithmic strain ${\mathbf{h}}^{e}$ and the logarithmic rate of transformation logarithmic  strain ${\mathbf{h}}^{tr}$ respectively,
\begin{equation}\label{eq:add_h_rate1}
\mathring{\mathbf{h}}^{e\_log}=\mathbf{D}^{e};~~\mathring{\mathbf{h}}^{tr\_log}=\mathbf{D}^{tr}
\end{equation}
Combine equation \ref{eq:add_D} and equation \ref{eq:add_h_rate1}, the following equation can be obtained.
\begin{equation}\label{eq:add_h_rate2}
\mathring{\mathbf{h}}^{log}=\mathring{\mathbf{h}}^{e\_log}+\mathring{\mathbf{h}}^{tr\_log}
\end{equation}
Apply logarithmic corotational integration\cite{khan1995continuum} in equation \ref{eq:add_h_rate2} on both sides, the following additive decomposition of total logarithmic strain can be received. Namely, the total logarithmic strain can be additively split into an elastic part corresponding to recoverable energy and transformation part associated with dissipated energy in transformation process. 
\begin{equation}\label{eq:add_h}
\mathbf{h}=\mathbf{h}^{e}+\mathbf{h}^{tr}
\end{equation}

\section{Model Formulation} \label{Model}
\subsection{General thermodynamic framework}

In order to develop finite strain SMAs model, we start with definition of the Gibbs free energy $ G $ to be a continuous function dependent on Kirchhoff stress tensor $ \bm{\uptau} $,  logarithmic strain of Eulerian type $ \mathbf{h} $, temperature $ T $, specific entropy $ s $, and a set of internal state variables $ \bm\Upsilon $ to be confirmed later on.
\begin{equation}\label{eq:GIBBS}
G(\bm{\uptau},\mathbf{h},T,s,\bm{\Upsilon}) = u - \dfrac{1}{\rho_{0}} \bm{\uptau} : \mathbf{h} - sT
\end{equation}
$ G $ is the Gibbs free energy, $\rho_{0}$ is the density at reference configuration, $ s $ is  specific entropy and $ u $ is internal energy. Later on, logarithmic transformation strain $ \mathbf{h}^{tr} $ and the martensitic volume fraction $ \xi $ will be chosen as internal state variables to model the SMAs nonlinear material response.

Based on the $ 2^{nd} $ law of thermodynamics, the dissipation energy $ \mathcal{D} $ can be written in the form of Clausius-Duhem inequality,
\begin{equation}\label{Dissipation}
\mathcal{D} = \mathbf{\bm{\sigma}:D} - \rho (\dot{u} - T\dot{s}) \geqslant 0
\end{equation}
$\rho$ is density at current configuration. Since Gibbs free energy is a continuous function as defined in equation \ref{eq:GIBBS}, take the logarithmic rate of equation \ref{eq:GIBBS}. Considering objective rates on a scalar variable equals to conventional time rate of that scalar, we are able to derive equation \ref{eq:GIBBS_LOG}. From now on, the log symbol in logarithmic rate will be ignored in later part for text legibility.
\begin{equation}\label{eq:GIBBS_LOG}
\mathring{G}^{log}= \dot{G} = \dot{u} - \dfrac{1}{\rho_{0}} \mathring{\bm{\uptau}}^{log} : \mathbf{h} - \dfrac{1}{\rho_{0}} \bm{\uptau} : \mathring{\mathbf{h}}^{log} -s\dot{T}-\dot{s}T
\end{equation}
 After some math on~\ref{eq:GIBBS_LOG}, we end up with equation \ref{eq:GIBBS_LOG_sb.} of the Gibbs free energy.
\begin{equation}\label{eq:GIBBS_LOG_sb.}
\dot{u} -\dot{s}T = \dot{G} + \dfrac{1}{\rho_{0}} \mathring{\bm{\uptau}}:\mathbf{h} 
+ \dfrac{1}{\rho_{0}} \bm{\uptau} : \mathring{\mathbf{h}} + s\dot{T}
\end{equation}
Substitute equation \ref{eq:GIBBS_LOG_sb.} into equation \ref{Dissipation}, the dissipation energy can be reformulated as follows,
\begin{equation}\label{eq:Dissipation_f}
\mathcal{D} = -\rho_{0}\dot{G}-\rho_{0}s\dot{T}-\mathring{\bm{\uptau}}:\mathbf{h} \geqslant 0
\end{equation}
Again, invoking Gibbs free energy is continuous, we are allowed to take chain rule differentiation on $G$ with respect to its independent variables. Noted $\bm{\uptau}$ and $\mathbf{h}$ are conjugate pair, and $s$ is also conjugated to $T$, only one from each of them is independent variable of Gibbs free energy $G$.
\begin{equation}\label{eq:Chain_Rule}
\dot{G}=\mathring{G}^{log} = \frac{\partial G}{\partial \bm{\uptau}}:\mathring{\bm{\uptau}}
+\frac{\partial G}{\partial T}\dot{T}+
\frac{\partial G}{\partial \bm{\Upsilon}}:\mathring{\bm{\Upsilon}}
\end{equation}
Substitute equation \ref{eq:Chain_Rule} into equation \ref{eq:Dissipation_f}, following equation for the dissipation energy $ \mathcal{D}  $ is derived :
\begin{equation}\label{eq:Dissipation_Cons}
\mathcal{D} = -(\rho_{0} \frac{\partial G}{\partial \bm{\uptau}} + \mathbf{h} ) :\mathring{\bm{\uptau}}
-(\rho_{0} \frac{\partial G}{\partial T} + s ) : \dot{T}
-\rho_{0}  \frac{\partial G}{\partial \bm{\Upsilon}} :\mathring{\bm{\Upsilon}} \geqslant 0
\end{equation}
Follow standard Coleman-Noll procedure, no matter what the thermodynamic path the system will have, the dissipation energy $ \mathcal{D} $ should always be greater than zero in order to satisfy the $2^{nd}$ thermodynamics law. the following constitutive relationship between conjugate pairs will be obtained.
\begin{equation}\label{eq:entropy_Cons}
s =- \rho_{0}\frac{\partial G}{\partial T}
\end{equation}
\begin{equation}\label{eq:h_Cons}
\mathbf{h} =- \rho_{0}\frac{\partial G}{\partial \bm{\uptau}}
\end{equation}
Apply constitutive equation \ref{eq:entropy_Cons} and equation \ref{eq:h_Cons} into dissipation energy inequality \ref{eq:Dissipation_Cons}, we have the following strict from of dissipation inequality.
\begin{equation}\label{Dissipation_State_V}
-\rho_{0}  \frac{\partial G}{\partial \bm{\Upsilon}} :\mathring{\bm{\Upsilon}} \geqslant 0
\end{equation}
\subsection{Constitutive modeling for SMAs at finite strain}
\subsubsection{Thermodynamic potential of constitutive model}
In this section, the general thermodynamic framework derived previously will be used to formulate the finite strain constitutive modeling for SMAs. This work is based on the infinitesimal strain model proposed by Lagoudas and coworkers \cite{boyd1996,lagoudas2008,lagoudas2012}, which has been extensively used for the design and development of SMA-based active device and smart structures \cite{peraza2013design,peraza2013opt,peraza2014origami} for the past two decades. We begin with an explicit expression for Gibbs free energy for start point. Independent variables of Gibbs free energy $G$ are chosen as kirchhoff stress $\bm{\uptau}$ and temperature $T$. Transformation logarithmic strain $\mathbf{h}^{tr}$ and martensitic volume fraction $ \xi $ are chosen as a set of internal state variables $ \mathbf{\Upsilon}=\{ \mathbf{h}^{tr},\xi\} $ to model the SMAs nonlinear material response. Transformation logarithmic strain $ \mathbf{h}^{tr} $ is accounting for the inelastic strain part caused by transformation between austenite and martensite phase, the martensite volume fraction $ \xi $  ranging from $ 0$ to $ 1 $ is used for differentiating the two different phases in SMAs, the explicit Gibbs free energy is given as:
\begin{equation}\label{eq:GIBBS_explicit}
\begin{aligned}
G(\bm{\uptau},T,\mathbf{h}^{tr}, \xi) =  -\dfrac{1}{2 \rho_{0}} \bm{\uptau} : \mathbf{S}:\bm{\uptau} - \dfrac{1}{\rho_{0}}  \bm{\uptau} :[~\bm{\alpha}(T-T_0)+\mathbf{h}^{tr}]\\+c \Big[(T-T_0)-T\ln (\dfrac{T}{T_0}) \Big]-s_0T+u_0+\dfrac{1}{\rho_{0}}f(\xi)
\end{aligned}
\end{equation}
$\mathbf{S}$ is the fourth-order compliance tensor dependent on martensitic volume fraction $\xi$, it is calculated by using a rule of mixtures as defined by equation \ref{eq:S_mix}, $ \bm{\alpha}$ is the second order thermoelastic expansion tensor, $\mathbf{h}^{tr}$ is transformation logarithmic strain, $ c $ is effective specific heat, $ s_0, u_0 $ are effective specific entropy at reference state and effective specific internal energy at reference state, respectively, they are defined similar as equation \ref{eq:S_mix} by virtue of rule of mixtures; $ T $ denotes current temperature while $ T_0 $ is reference temperature. $ f(\xi) $ is a transformation hardening function upon being defined later on.
\begin{equation}\label{eq:S_mix}
\mathbf{S}(\xi)=\mathbf{S}^A + \xi(\mathbf{S}^M-\mathbf{S}^A)=\mathbf{S}^A + \xi\Delta\mathbf{S}
\end{equation}
Using rule of mixture to calculate the effective compliance tensor in equation \ref{eq:S_mix}, $\mathbf{S}^A$ is forth order compliance tensor for austenite phase, $\mathbf{S}^B$ is forth order compliance tensor for martensite phase, and $\Delta\mathbf{S}$ is the difference between them.

From Lagoudas infinitesimal model\cite{boyd1996,lagoudas2008,lagoudas2012}, we take the same hardening function as defined by equation \ref{eq:hardening}, in which $ b^A,b^M,\mu_1,\mu_2 $ are material parameters defined in equation \ref{Para_Y}.
\begin{equation}\label{eq:hardening}
f(\xi)=\begin{cases} \dfrac{1}{2} \rho_0 b^M {\xi}^2+(\mu _1+\mu_2)\xi, \; \dot{\xi}>0, \vspace{5pt} \\ \dfrac{1}{2} \rho_0 b^A {\xi}^2+(\mu _1-\mu_2)\xi, \; \dot{\xi}<0, \end{cases}\\
\end{equation}
To this end, Gibbs free energy $G$ is explicitly defined in equation \ref{eq:GIBBS_explicit}. The next step is to apply it to general thermodynamic framework to obtain constitutive equations. Substitute Gibbs free energy expression into equation \ref{eq:entropy_Cons} to obtain the constitutive relation for entropy $s$, and also substitute into equation \ref{eq:h_Cons} to obtain the constitutive relation for logarithmic strain $\mathbf{h}$.
\begin{equation}\label{eq:entropy_Cons_f}
s =- \rho_{0}\frac{\partial G}{\partial T}=\dfrac{1}{\rho_{0}} \bm{\uptau}:\bm\alpha+c\ln (\dfrac{T}{T_0}) + s_0
\end{equation}
\begin{equation}\label{eq:h_Cons_f}
\mathbf{h} = - \rho_{0}\frac{\partial G}{\partial \bm\uptau}=\mathbf{S~\bm\uptau}+\bm\alpha(T-T_0)+ \mathbf{h}^{tr}
\end{equation}   
Transformation logarithmic strain $ \mathbf{h}^{tr} $ and martensite volume fraction $ \xi $ are internal state variables to model the nonlinear system. Rewrite the strict from dissipation inequality equation (\ref{Dissipation_State_V}) by choosing the set of internal state variables $ \mathbf{\Upsilon}=\{ \mathbf{h}^{tr},\xi\} $
\begin{equation}\label{eq:Dissipation_State_V2}
-\rho_{0}  \frac{\partial G}{\partial \mathbf{h}^{tr}} :\mathring{\mathbf{h}}^{tr} -\rho_{0}  \frac{\partial G}{\partial \xi}\dot{\xi} \geqslant 0
\end{equation}
\subsubsection{Evolution equation of internal state variables}\label{sec:Evolution}
In the subsection, we will set up the evolution equation between transformation logarithmic strain $ \mathbf{h}^{tr} $ and martensite volume fraction $ \xi $. In this model, only detwinned martensite variant is considered in the transformation process. One key assumption from Lagoudas infinitesimal model is: any change in the current microstructural state of the material is strictly a result of a change in the martensitic volume fraction (Boyd and Lagoudas\cite{boyd1996,lagoudas2008}). The rigorous mathematical derivation of this assumption is provided by Qidwai and Lagoudas \cite{qidwai2000}, in which they used the principle of maximum dissipation and the $J_2$ plasticity theory to derive that the remaining internal state variables in strict form of dissipation inequality \ref{eq:Dissipation_State_V2} is directly proportional to the evolution of martensitic volume fraction $\xi$. Inspired from that, we thus propose the following evolution relationship between $ \mathbf{h}^{tr} $ and   $\xi$.
\begin{equation}\label{eq:evolution}
{\mathring{\mathbf h}}^{tr}= \bm{\Lambda}  \dot{\xi},  \ \  \bm\Lambda=\begin{cases}\bm{\Lambda}^{fwd}, \; \dot{\xi}>0, \vspace{5pt} \\ \bm{\Lambda}^{rev}, \; \dot{\xi}<0, \end{cases}\\
\end{equation}
where, $\bm\Lambda^{fwd}$ is the transformation direction tensor during forward transformation process, while $\bm\Lambda^{rec}$ is the transformation direction tensor during reverse transformation process , they are defined as following:
\begin{equation}\label{eq:direction}
\bm\Lambda^{fwd}=
\frac{3}{2} H^{max} 
\frac{\bm{\uptau}^{'}}{\bar{\uptau}^{'}},  \ \bm\Lambda^{rev}=
H^{max} \frac{\mathbf h^{tr-r}}{{\xi}^{r}}.
\end{equation}
In which, $H^{max}$ is a material parameter denoting the maximum transformation strain. In forward transformation direction tensor, $ \bm{\uptau}^{'} $ is the deviatoric stress tensor calculated by {\small $ \bm{\uptau}^{'} =\bm{\uptau} -{\small \frac{1}{3}}\textrm{tr}(\bm{\uptau})~\mathbf{I} $}, where $ \mathbf{I} $ is the second order identity tensor. The effective (von Mises equivalent) stress is given by $ \bar{\uptau}^{'} ={\small \sqrt{{{\small \frac{3}{2}}\bm\uptau}^{'}:\bm{\uptau}^{'}}}$.  In reverse transformation direction tensor, $\mathbf h^{tr-r}$ represents the transformation logarithmic strain at the reverse transformation starting point; ${\xi}^{r}$ denotes the martensitic volume fraction at the reverse transformation starting point. The definition of transformation tensor is based on the assumption that transformation strain will evolve following the direction of deviatoric stress during forward $ (\dot{\xi}>0) $. During the reverse transformation $(\dot{\xi}<0)$, the transformation strain will decrease proportionally from the value at reversal starting point to the finish value of reverse transformation.

\subsubsection{Transformation function}\label{Trans_Func}
After define the evolution equation between the transformation logarithmic strain $ \mathbf{h}^{tr} $ and the martensitic volume fraction $\xi$ in section \ref{sec:Evolution}, the next objective is to define a proper transformation function as criteria to determine when the transformation will happen. Substitute the evolution equation \ref{eq:evolution} into the dissipation inequality (\ref{eq:Dissipation_State_V2}), we obtain the following equation:
\begin{equation}\label{eq:Dissipation_xi}
(\bm\uptau:\bm\Lambda-\rho_{0}  \frac{\partial G}{\partial \xi})\dot{\xi}=\pi\dot{\xi}\geqslant 0 
\end{equation}
Where, the scalar quantity $\pi$ is called general thermodynamic driving force conjugated to $\xi$. Substitute the specific Gibbs free energy expression equation \ref{eq:GIBBS_explicit} into equation \ref{eq:Dissipation_xi}, the expression for general thermodynamic driving force $ \pi $ is given by:
\begin{equation}\label{eq:Driving_Force}
\begin{aligned}
\pi(\bm\uptau,T,\xi)=\bm\uptau:\bm\Lambda+
\dfrac{1}{2}\bm\uptau:{\Delta}\mathbf{S}:\bm\uptau+\bm\uptau:{\Delta}\mathbf{\alpha}(T-T_0)-\rho_0\Delta c \\ 
\big[ T-T_0-T\ln(\dfrac{T}{T_0}) \big ] + \rho_0\Delta s_0 T - \rho_0\Delta u_0 - \frac{\partial f}{\partial \xi}
\end{aligned}
\end{equation}
Where, material parameters $\Delta \mathbf{\alpha}, \Delta c, \Delta s_0, \Delta u_0 $ represents the difference for different phases. They have similar definition as $\Delta \mathbf{S}$ in equation (\ref{eq:S_mix}). From Lagoudas model, we are also assuming that whenever the thermodynamic driving force $\pi$ reach a critical value $Y $ (or $ -Y $), the forward (reverse) phase transformation will take place. Thus we introduce a transformation function $\Phi$ defined by (\ref{eq:Transfor_Fun}) as the criteria to determine whether forward and reverse transformation happens. 
\begin{equation}\label{eq:Transfor_Fun}
\normalfont{\Phi}=\begin{cases}~~\pi - Y, \; \dot{\xi}>0, \vspace{5pt} \\ -\pi - Y, \; \dot{\xi}<0, \end{cases}\\
\end{equation}
It is proved from Qidwai and Lagoudas\cite{qidwai2000} that some certain constraints have to be applied on the evolution of martensitic volume fraction $\xi$, this constraint can be expressed as so-called Kuhn-Tucker conditions:
\begin{equation}\label{eq:K-T}
\begin{aligned}
\dot{\xi} \geqslant 0; \quad \Phi(\bm\uptau,T,\xi)= ~~\pi - Y \leqslant 0;  \quad  \Phi\dot{\xi}=0;~~~\bf{(A\Rightarrow M)}\\
\dot{\xi} \leqslant 0; \quad \Phi(\bm\uptau,T,\xi)= -\pi - Y \leqslant 0; \quad   \Phi\dot{\xi}=0;~~~ \bf{(M\Rightarrow A)}
\end{aligned}
\end{equation}
The critical value $Y$ together with the other material parameters $ (b^A,b^M,\mu_1,\mu_2) $ defined in the hardening function (\ref{eq:hardening}) are determined from the phase diagram parameters $(M_s, M_f , A_s, A_f, \Delta s_0, \Delta u_0)$. The readers themselves are encouraged to find calibration details through the book authored by Lagoudas and coworskers \cite{lagoudas2008}.
\begin{equation}\label{Para_Y}
\begin{aligned}
\begin{cases} Y=\tfrac{1}{4}\rho_0\Delta s_0(M_s+M_f-A_s-A_f) \\
b^A=-\Delta s_0(A_f-A_s)\\
b^M=-\Delta s_0(M_s-M_f)\\
\mu_1=\tfrac{1}{2}\rho_0\Delta s_0(M_s+M_f)-\rho_0\Delta u_0\\
\mu_2=\tfrac{1}{4}\rho_0\Delta s_0(A_s-A_f-M_f+M_s)
\end{cases}
\end{aligned}
\end{equation}

\section{Numerical Results}
\label{Result}
In this section, several boundary value problems, such as a simple SMAs bar under actuation loading, a SMAs beam under bending loading, and a SMAs tube under torsion loading are gonging to be analyzed to test the capability of the proposed finite strain SMAs model. The results, such as actuation loading curve and accumulated residual stress, predicted by proposed finite strain model and its infinitesimal counterpart will be compared. 

\subsection{Bar Actuation Problem}
The first boundary value problem analyzed is a simple SMAs bar under actuation loading. The bar length is $L=100$ mm, the cross section is a square with its edge length $ a=10 $mm.The boundary conditions are left face fixed in all degrees of freedom and the right face subject to constant pressure. 

\begin{figure}[H]
	\centering
	\includegraphics[width=0.5\textwidth]{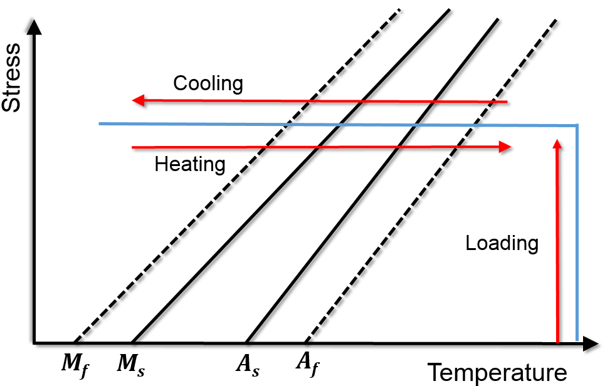}
	\caption{Loading path for bar actuation problem}
	\label{fig:BarPath}
\end{figure}

%\begin{figure}[H]
%	\centering
%	\includegraphics[width=0.6\textwidth]{figures/Bar_Actuation.png}
%	\caption{Schematic for bar actuation problem }
%	\label{fig:Bar}
%\end{figure}
%
As depicted in loading diagram \ref{fig:BarPath}, the bar is first loaded up to constant pressure 30 MPa. Then, the bar is cool down from the initial temperature 360$^{\circ}$C to lower temperature 220$^{\circ}$C. The SMA bar will experience a large extension due to its forward transformation from austenite phase to detwinned martensite phase. The next step is increasing temperature from 220$^{\circ}$C to 360$^{\circ}$C, the bar will contract to its original length due to the reverse transformation from detwinned martensite phase to austenite phase. This SMAs bar experienced a full actuation loading cycle. The material parameters utilized in this SMAs bar analysis is from table \ref{table:MP}\cite{lagoudas2008}. 

\begin{table}[H]
	\centering
	\caption{Material parameters of SMAs}
	\label{table:MP}
	\begin{tabular}{@{}lccccccccc@{}}
		\toprule
		&E\textsubscript{A}& E\textsubscript{M}   & M\textsubscript{s}  &M\textsubscript{f} &A\textsubscript{s} &A\textsubscript{f}\\ 
		&90 GPa &47 GPa &308$^{\circ}$C &246$^{\circ}$C &284$^{\circ}$C &356$^{\circ}$C\\ \midrule
		&$\alpha_A$ &$\alpha_M$ &$\textit{v}_A$ &$\textit{v}_M$\\ 
		&2.2e-5  & 2.2e-5 & 0.33  &0.33\\ 
		\bottomrule
	\end{tabular}
\end{table}

\begin{figure}[H]
	\centering
	\centering
	\includegraphics[width=0.5\textwidth]{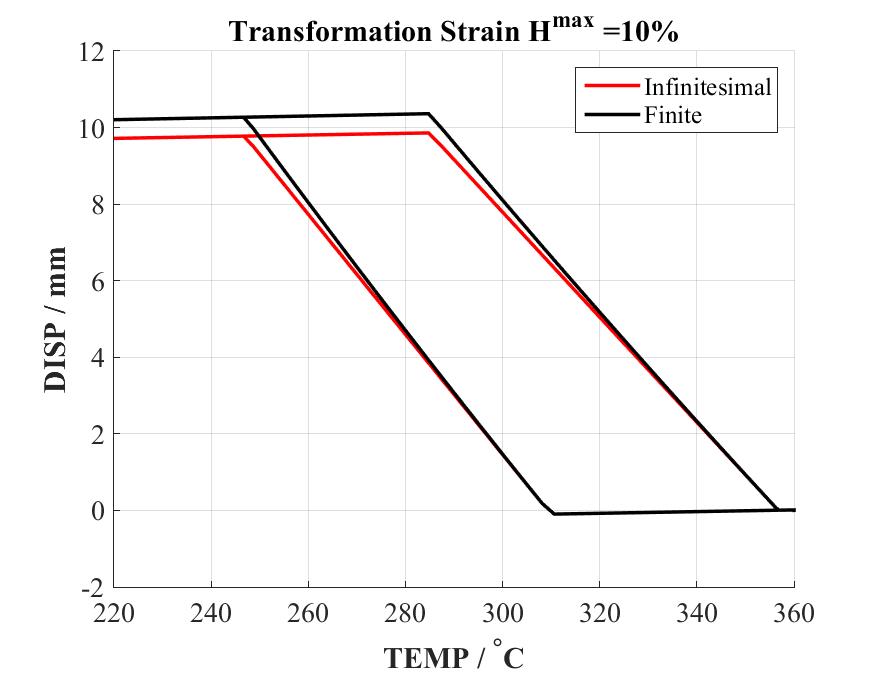}
	\caption{ $ {H}^{max}=10\%$ }
	\label{fig:H10}
\end{figure}

In this simple SMAs bar actuation problem, the results predicted from both proposed finite strain and infinitesimal strain model are compared in figure \ref{fig:H10}, which are plotted by temperature versus displacement. As the results showing in figure \ref{fig:H10}, with the  material parameter maximum transformation strain ${H}^{max}=10\%$, the predicted results between the infinitesimal model and the proposed finite strain model is slightly different.

%This indicates it is more appropriate to use a finite strain model to predict SMAs material response with large strains.

%\begin{figure}
%	\centering
%	\begin{minipage}{.5\textwidth}
%		\centering
%		\includegraphics[width=1\textwidth]{figures/H10.jpg}
%		\captionof{figure}{Comparison of results for ${H}^{max}=10\% $}
%		\label{fig:H10}
%	\end{minipage}%
%	\begin{minipage}{.5\textwidth}
%		\centering
%		\includegraphics[width=1\textwidth]{figures/H15.jpg}
%		\captionof{figure}{Comparison of results for ${H}^{max}=15\% $}
%		\label{fig:H15}
%	\end{minipage}
%\end{figure}

\subsection{Beam Bending Problem}
\begin{figure}[H]
	\centering
	\includegraphics[width=0.6\textwidth]{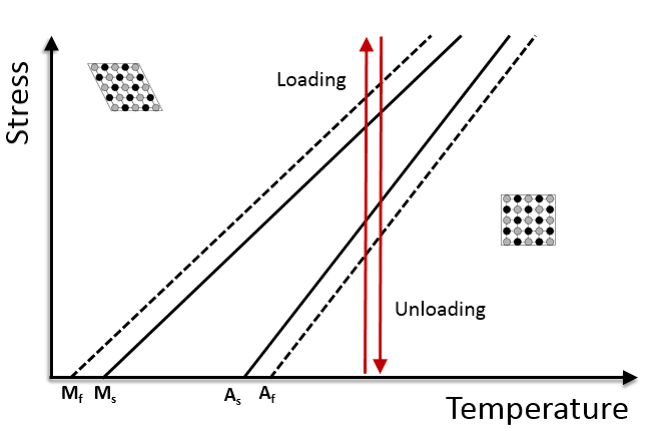}
	\caption{Loading path for beam bending problem}
	\label{fig:Beam_path}
\end{figure}

The second problem analyzed is a three dimensional SMA beam under bending. The material parameters used are also from table \ref{table:MP}. Boundary conditions are left side fixed in all degrees of freedom and the right side free to move at all directions. The loading path is described as figure \ref{fig:Beam_path}. The beam starts from austenite phase under bending pressure which is gradually increasing from zero to maximum value. The SMAs beam undergoes a forward stress-induced phase transformation from austnite phase to detwinned martensite phase. Then, the pressure gradually decreases from maximum to zero, during which the beam experiences a reverse transformation from detwinned martensite phase to austenite phase. 

%\begin{figure}[H]
%	\centering
%	\includegraphics[width=0.6\textwidth]{figures/Beam_Bending.png}
%	\caption{Schematic for beam bending problem}
%	\label{fig:Beam}
%\end{figure}

During this loading cycle, phase transformation in SMAs is the only fact accounting for inelastic material response. So once it is fully recovered from the transformation, the SMAs beam should be able to return to unstress state. However, different numerical results are observed between finite strain SMAs model and its infinitesimal counterpart.   

\begin{figure}[H]
	\centering
	\includegraphics[width=0.6\textwidth]{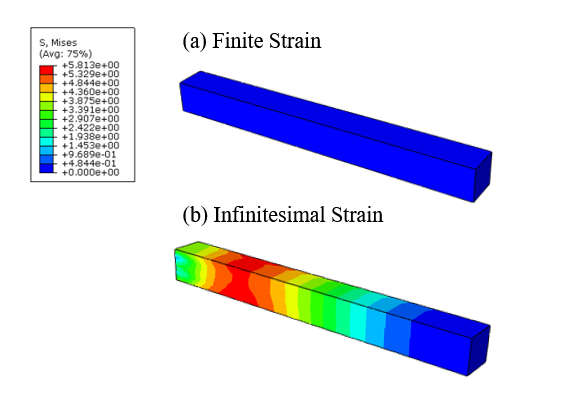}
	\caption{Accumulated residual von mises stress of beam after bending}
	\label{fig:Residual_beam}
\end{figure}

As irt shows in figure \ref{fig:Residual_beam}(a), residual von mises stress results given by finite strain SMAs model has magnitude of $10^{-5}$ in the end. As a comparison, residual von mises stress results predicted by infinitesimal strain model is around a nontrivial 5 MPa at the end of the loading. This comparison shows for SMAs components with large rotations, response predicted by infinitesimal strain model will accumulate spurious residual stress at the end of the loading, while this can be eliminated by using a proposed finite strain model. This demonstrates a consistent finite strain model is needed to provide more accurate results.  

\subsection{Torque Tube Problem}
\begin{figure}[H]
	\centering
	\includegraphics[width=0.6\textwidth]{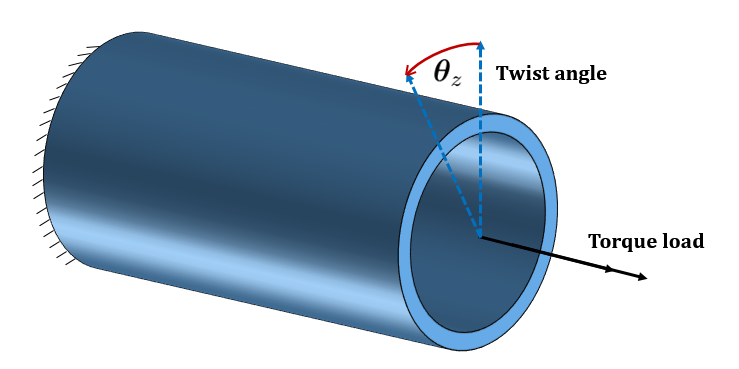}
	\caption{Schematic for torque tube problem}
	\label{fig:tube}
\end{figure}

Boundary value problem of hollow cylindrical torque tube under torsion loading is solved in this section. In this problem, apart from the large shear strain SMAs will exhibit, it also undertakes large geometry rotations. The boundary conditions are described in figure \ref{fig:tube}, The tube left face is fixed in all degrees of freedom and tube right face is subject to twist $\theta_z(t)$ angle along z axis. The material parameters are also choosing from table \ref{table:MP}, the maximum transformation strain is  ${H}^{max}=5\% $ in this case.  

\begin{figure}[H]
	\centering
	\includegraphics[width=0.6\textwidth]{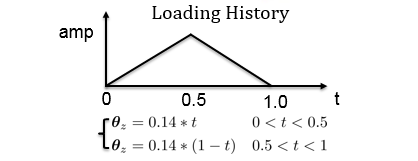}
	\caption{Torque tube loading history}
	\label{fig:tube_history}
\end{figure}

The loading history for troque tube is in figure \ref{fig:tube_history}. The tube starts with zero twist angle in austenite phase. Then, the twist angle is gradually increasing  proportionally from zero to a maximum value. After twist angle reaches the peak amplitude, it gradually decreases linearly from peak to the zero value at the end of loading. The SMAs tube will experience a full pseudoelastic loading cycle.  

%As showing in figure \ref{fig:tube_results}, the results are plotted by shear stress versus twice of the shear strain. Results predicted by proposed finite strain SMAs model are compared to Lagoudas infinitesimal counterpart. It should be noted that the infinitesimal strain model itself can not predict SMAs response with large rotations, the Abaqus nonlinear geometry solver (NLGEOM) with rotation procedure makes it able to capture nonlinear geometry rotations. From figure \ref{fig:tube_results}, the comparison shows SMAs pseudoelastic response predicted by proposed finite strain model is slightly softer than the results provided by infinitesimal strain model with NLGEOM solver.
%  
%\begin{figure}[H]
%	\centering
%	\includegraphics[width=0.8\textwidth]{figures/Tube_Results.jpg}
%	\caption{Shear stress versus shear strain of torque tube under pseudoelastic loading}
%	\label{fig:tube_results}
%\end{figure}

\begin{figure}[H]
	\centering
	\includegraphics[width=0.9\textwidth]{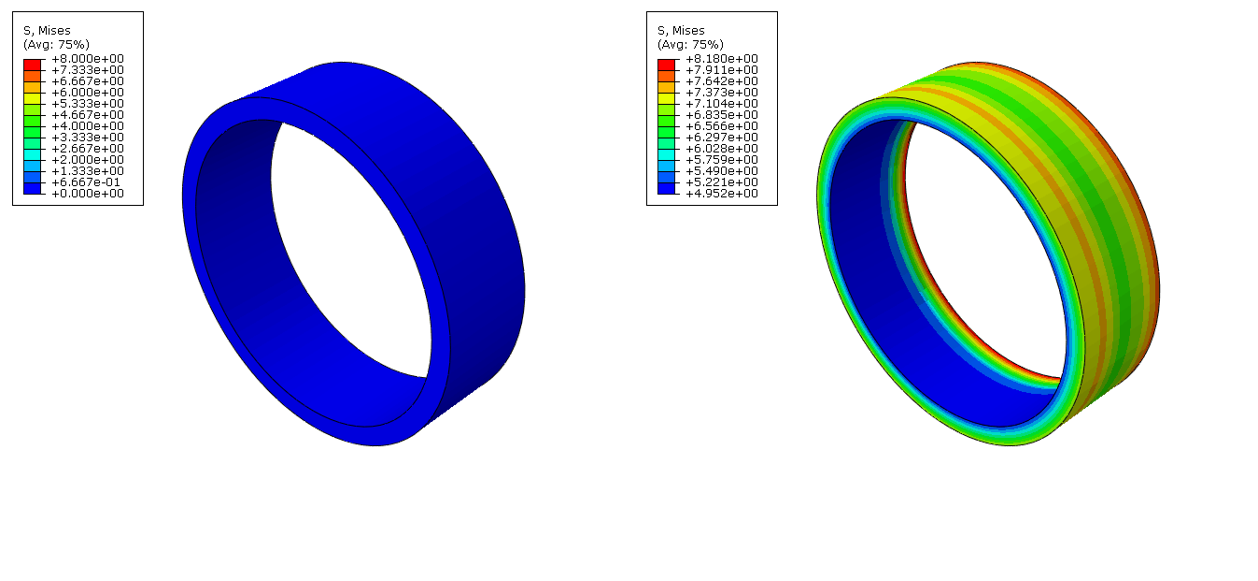}
	\caption{Accumulated residual von mises stress for torque tube after torsion}
	\label{fig:Residual_tube}
\end{figure}

 Difference on predicted results are observed on accumulated residual stress at the end of loading cycle. Since transformation is the only inelastic process considered in SMAs for current analysis, torque tube should return to unstressed state once it fully recover from the torsion loading. As showed on the left picture in figure \ref{fig:Residual_tube}, this was correctly simulated by the proposed finite strain model. However, it is not the case for the results received from infinitesimal strain model. The residual stress remain around 4 MPa at the end of loading as showing in right picture of figure \ref{fig:Residual_tube}. It is discussed in the introduction part, the residual stress is coming from the non-integral issue of many objective rates have.  As spurious residual stress accumulated through a number of loading cycles, it may even triggers the transformation to happen which should not be the real case in reality.  Using the consistent finite strain model based on logarithmic strain and logarithmic rate will resolve this problem. Again, the residual stress results in torque tube problem demonstrate the necessity of utilizing the proposed finite strain model to predict the SMAs response under large deformations. 

%%==========================================================================

\section{Conclusions}\label{Conc}
Based on the constitutive model for shape memory alloys within small deformation range proposed by Lagoudas and coworkers\cite{boyd1998,lagoudas2008,lagoudas2012}, a three dimensional constitutive model for martensitic transformation in shape memory alloys is formulated based on additive decomposition of the rate of deformation tensor. The model is derived fully consistently within thermodynamic framework by utilizing the logarithmic strain as finite strain measure, whose logarithmic rate is equivalent to the rate of deformation tensor. The martensitic volume fraction and the second-order transformation logarithmic strain tensor are chosen as the internal state variables accounting for the inelastic transformation process. Numerical simulations considering basic SMA component geometries such as a bar, a beam and a torque tube are performed to test the capabilities of proposed model under both mechanically and thermally induced phase transformation. For numerical examples in simple bar with maximum transformation strain $ {H^{max}=10\%} $, discrepancies are observed between the responses predicted by the proposed model and its infinitesimal counterpart. Also, spurious residual stress observed in predicted results by infinitesimal strain model are successfully eliminated by proposed model. This shows that the infinitesimal strain assumption is not applicable when SMA components exhibit finite strains along with finite rotations, the proposed consistent finite strain model based on logarithmic strain and logarithmic rate is needed to give more accurate results. The presented formulation will be extended in future work for considering transformation-induced plasticity in cyclic loadings.

\section{Acknowledgments}\label{Ack}
The author is deeply grateful for the financial support provided by the Qatar National Research Fund under Grant NPRP 7-032-2-016. 

\bibliographystyle{aiaa} 
\bibliography{myarticle}

\end{document}